**Large-Volume Intrathecal Administrations: Impact on CSF Pressure and Safety Implications**

Vasily Belov[1-3*], Janine Appleton[1], Stephan Levin[1], Pilar Giffenig[1], Beata Durcanova[1], Mikhail Papisov[1-3]

[1]Massachusetts General Hospital, Boston, MA, United States
[2]Harvard Medical School, Boston, MA, United States
[3]Shriners Hospitals for Children-Boston, Boston, MA, United States

**\* Correspondence:**

Vasily Belov
vbelov@mgh.harvard.edu

## ABSTRACT

The increasing number of studies demonstrates the high potency of the intrathecal (IT) route for the delivery of biopharmaceuticals to the central nervous system (CNS). Our earlier data exhibited that both the infused volume and the infusion rate can regulate the initial disposition of the administered solute within the cerebrospinal fluid (CSF). This disposition is one of key factors in defining the subsequent transport of the solute to its intended target. On the other hand, fast additions of large volumes of liquid to the CSF inevitably raise the CSF pressure (a.k.a. intracranial pressure (ICP)), which may in turn lead to adverse reactions if the physiologically delimited threshold is exceeded. While long-term biological effects of elevated CSF pressure (hydrocephalus) are known, the safety thresholds pertaining to short-term ICP elevations caused by IT administrations have not yet been characterized. This study aimed to investigate the dynamics of ICP in rats and non-human primates (NHP) with respect to IT infusion rates and volumes. The safety regimes were estimated and analyzed across species to facilitate the development of translational large-volume IT therapies. The data revealed that the addition of a liquid to the CSF raised the ICP in a rate and volume-dependent manner. At low infusion rates (<0.12 ml/min in rats and <2 ml/min in NHP), NHP and rats displayed similar tolerance patterns. Specifically, safe accommodations of such added volumes were mainly facilitated by the accelerated pressure-dependent CSF drainage into the blood, with ICPs stabilizing at different levels below the safety threshold of 28±4 mm Hg in rats and 50±5 mm Hg in NHPs. These ICPs were safely tolerated for extended durations (of at least 2-25 min). High infusion rates (including boluses) caused uncompensated exponential ICP elevations rapidly exceeding the safety thresholds. Their tolerance was species-dependent and was facilitated by the compensatory role of the varied components of craniospinal compliance while not excluding the possibility of other contributing factors such as lymphatic drainage specifically in rats. In conclusion, large volumes of liquids can safely be delivered via IT routes provided that ICP is monitored as a safety factor and cross-species physiological differences accounted for.

## INTRODUCTION

The IT drug delivery route is of increasing interest due to the absence of continuous barriers between the CSF and CNS[1], which makes the fraction of CSF-borne compounds delivered to the brain from the CSF through perivascular channels[2,3,4] pharmacologically significant[5,6,7,8,9,10]. This is especially important for macromolecules, including biopharmaceuticals, for which the access to CNS from the systemic circulation is particularly hindered due to large molecular size and the presence of several vascular barriers (blood-brain, blood-arachnoid, and blood-CSF).[11,12]

In the clinical setting, the most commonly used routes for accessing the CSF are: (1) the intrathecal lumbar (ITL) route through the lumbar cistern (enveloping the cauda equina) and (2) the more invasive intracerebroventricular (ICV) route through the brain parenchyma into one of the lateral ventricles of the brain. Although the ICV route is clinically feasible, it requires the surgical installation of a transcranial port with an accompanying cannula (Omaya reservoir). The ITL route can be utilized for chronic use (with surgical installation of subcutaneous ports or electronic pumps) as well as for one-dose injections directly through vertebral disks in the L3-L5 region. Both methods are relatively simple, well developed, and particularly suitable for biopharmaceuticals, such as gene vectors, intended for single injections or multiple injections given at a large interval (several months).

The ITL route is commonly used in pain and spasticity management for chronic delivery of respective drugs from subcutaneously implanted electronic pumps. Such pumps deliver very small volumes directly to the target (nerve roots). The action of the drug delivered in such a way is focal in nature in that it does not spread cranially along the spinal column. Our data obtained earlier in non-invasive PET studies showed that such local action was made possible due to the absence of directional CSF flow within the spine.[5,13] We also found that the ITL administration of a larger volume caused immediate translocation of the administered solute to the cerebro-cervical area, resulting in a nearly identical initial disposition in the CSF as compared to ICV administration.[7] Rieselbach et al.[14] demonstrated that both humans and non-human primates could tolerate large-volume ITL boluses of up to 33% and 42% of the total CSF volume, respectively. Therefore, the ITL route is currently of significant interest for both testing the efficacy of novel biopharmaceuticals in preclinical models as well as for clinical use. However, the physiological mechanisms responsible for the accommodation of such additional large volumes by the subarachnoid (leptomeningeal) space as well as the safety limits of different IT administration regimes remain unknown.

Due to the confinement of the subarachnoid liquid-filled space by rigid (bone) boundaries, the modified Monroe-Kellie doctrine[15] states that the total volume of its contents must remain constant. The brain, CSF, and blood are the major constituents, the volume fractions of which under normal conditions are 83%, 11%, and 6%, respectively. Given the constant volume of the brain tissue, it is the hydrodynamic balance of blood and CSF volumes that primarily determines the CSF pressure (often referred to as intracranial pressure, or ICP). In this respect, experimentally measurable characteristics of ICP can be evaluated as safety factors characterizing the impact of IT administrations.

ICP is not constant and on a long-term scale (hours), depends on the balance of CSF production and drainage. On a shorter-term scale (seconds), ICP fluctuates due to the rhythmic changes of the cerebral blood volume. Such a dynamic nature of the ICP signal necessitates analysis of not only the mean value, but also of the ICP waveform[16,17]. The latter is characterized by periodic pulses each consisting of three notches P1, P2, and P3, progressively decreasing in amplitude and reflecting propagation of the arterial pulse pressure wave. P1 harmonic is a fundamental component (percussion wave) that represents transmission of the arterial pulse through the choroid plexus to the CSF. The amplitude of this component and the average ICP are the key parameters used for characterizing the ICP autoregulatory capacity i.e. the ability to maintain a normal ICP. This capacity can be assessed by studying the ICP-CSF volume and the ICP amplitude-mean ICP relationships. There is significant evidence indicating the successful use of these relationships for deriving indices effective for the diagnosis and assessment of treatments for neurological disorders affecting the liquid homeostasis in the CNS such as subarachnoid and intracerebral hemorrhage, ischemic stroke, hydrocephalus, meningitis/encephalitis, and traumatic brain injury among others.[16,17,18,19]

In humans, 7-15 mm Hg is considered a normal range of ICP as measured via lumbar access in supine adults.[20] While ICP in humans is commonly reported for the supine position, the measured value is known to be a function of a body posture.[17] For example, in the vertical position, ICP baseline is negative with a mean of around -10 mm Hg, but not exceeding -15 mm Hg.[21] Although tolerable, persistently elevated ICP levels of 15-22 mm Hg are considered hypertensive and are indicative of an underlying pathology requiring treatment.[16] Aggressive therapy is typically initiated when ICPs surpass 22 mm Hg.[17,22] These pressures are characteristic of impaired autoregulatory capacity and if present for sustained periods of time (>37 min in adults and >8 min in children[23]), may be predictive of poor patient outcomes in regards to survival rates and a long-term functionality.[17] Moreover, the modeled relationship between elongated periods of high ICPs and poor outcomes was found to be nearly exponential in both the adult and pediatric populations.[23] ICP levels exceeding 40 mm Hg are characteristic of acute brain injuries and cannot be tolerated for extended durations (>30 min).[16,24] It should be noted that reported ICP thresholds are even lower for special populations such as the pediatric, elderly, and female populations.[23,25] One of the major factors associated with adverse events of elevated ICP and its duration is the cerebral perfusion flow, critical decrease of which can cause secondary ischemic injury.[16] Cerebral perfusion flow is regulated by the cerebral perfusion pressure (CPP) defined by the difference between mean arterial pressure and ICP.[26,27] The current guidelines recommend maintaining CPP between 60 and 70 mm Hg.[22] This approach aims to prevent secondary injuries due to hypoperfusion (e.g., cerebral ischemia) or hyperperfusion (e.g., edema).[16] Increased ICP levels also adversely impact the brainstem function as excessive pressures can cause bradycardia and hypertension (Cushing reflex), and if left untreated, precipitate respiratory depression and death.[28]

Hypothetically, the extent of adversity to large-volume IT injections/infusions should depend not only on the magnitude of the CSF pressure elevation but also the duration of such increases. This fact implies that both the volume and the delivery rate of liquids added to the CSF contribute to the safety thresholds provided that the relationship between these factors and the CSF pressure is established. These thresholds, least the physiological mechanisms underlying them, have not been studied systematically with respect to IT administrations, in neither humans nor laboratory animals.

Considering prior advances in ICP research obtained primarily from trauma patients, this study aims to establish a method of ICP monitoring for characterizing disbalances in the CSF pressure caused by various modes of IT administrations. The safety thresholds are estimated and analyzed across species to provide recommendations for translational studies aiming at developing clinical approaches for monitoring the safety of IT therapy delivery. To this end, we carried out experiments in rats and NHPs and: (1) determined the relationship between ICP, injection rates and volumes, (2) characterized the safety thresholds for ICP elevations, and (3) investigated the patterns of ICP relaxation to the baseline levels after bolus injections. The obtained data are discussed within the context of current knowledge on the involved physiological mechanisms and translational implications.

## MATERIALS AND METHODS

### Animals

All animal studies were carried out in accordance with the protocols approved by the Institutional Animal Care and Use Committee of Massachusetts General Hospital. Male Sprague-Dawley CD rats (n=8, 325±76 g) were obtained from Charles River Laboratories (Shrewsbury, MA). One cynomolgus (*Macaca Fascicularis*, 7.5 kg) and one rhesus monkey (*Macaca Mulatta*, 8.2 kg) were obtained from Northern Biomedical and were equipped with subcutaneous injection ports (P.A.S. Port *Elite*, Smiths Medical ASD, Inc.). The ports were connected to polyurethane kink-resistant catheters (Smiths Medical ASD, Inc.) entering the subarachnoid space at the L4/L5 spinal segment and were advanced to the L1/T12 area. Both ports were surgically implanted more than 3 months prior to the study and maintained as recommended per the manufacturer. Catheter potency and absence of leaks were confirmed by PET imaging with macromolecular tracers labeled with [89]Zr, as in our previous studies.[5,6,7] By the time of these studies, the ports had one-way (inwards) potency, likely due to the development of tissue flaps around the catheter openings. All animals were kept on a 12 hours/12 hours light/dark cycle, switching on at 7 AM and off at 7 PM. Food and water were provided ad libitum to rats. Monkeys had three meals a day while the water was provided ad libitum.

### ICP registration methodology

ICP in rats was measured directly in the cisterna magna pool of the CSF using a 1F piezoresistive diffused semiconductor pressure sensor mounted on the tip of a 20 cm long 0.8 F polyimide catheter (Millar, Inc). The pressure catheter was inserted through a temporary angiocatheter installed through the atlanto-occipital membrane. In monkeys, port configuration did not allow sensor insertion into the leptomeningeal space. The pressure ($ICP_L$) was measured externally, using the same sensor inserted into a saline-filled catheter line connected to the subcutaneous IT port. Electrical signals from the sensor were processed using a FE221 Bridge Amplifier (ADInstruments) coupled with a PowerLab 4/35 data acquisition platform (ADInstruments). Control of both modules, signal acquisition, and subsequent ICP waveform analysis were performed using a LabChart 8 software (ADInstruments) installed on an Apple MacBook Pro laptop. ICP sampling rate was set to 1000 measurements per second. Prior to all ICP measurements, the pressure transducer was soaked in sterile water or saline at room temperature for at least 30 min. After that, atmospheric pressure was set to 0 mm Hg and the stability of the sensor signal was ensured by recording the baseline for at least 10 min.

### Rats

#### Cisterna magna catheterization

Percutaneous non-surgical catheterization of the cisterna magna was carried out following the technique developed by Jeffers and Griffith.[29] Animals were anesthetized by inhalation of isoflurane/air mixture (3% for induction, 2% for maintenance) given at 300 ml/min flow rate using a SomnoSuite® small animal anesthesia platform (Kent Scientific). The animal was then

                                                       

mounted on a stand inclined to 50° angle with the head flexed down, thereby positioning the occipital bone in the horizontal plane. The back of the neck and the base of the skull were shaved and disinfected with 70% ethyl alcohol. The location of the cisterna magna was identified by palpation of a 3-mm$^2$ rhomboid depressed area (atlanto-occipital joint) in the middle of the line connecting the ear bases. A 24G 0.75" IV catheter (Angiocath™, Becton Dickinson) was then slowly inserted to a depth of 7 mm. The inner needle was immediately removed after the appearance of the clear CSF in the outer catheter. The catheter's Luer adapter was then promptly capped, and the catheter was affixed to the skin with a cyanoacrylate gel. Localization of the tip of the (radiopaque) catheter in the cisterna magna was verified at the end of the study by taking a CT scan of a rat's head (Fig. S1A).

## ICP measurements

ICP measurements in rats were carried out using the equipment configuration shown in Fig. S1B. Specifically, after the cisterna magna catheterization, a Tuohy Borst adapter (Cook Medical) equipped with a side arm (initially closed) and pre-filled with saline was connected to the IV catheter via a Luer adapter. A micro-tip pressure transducer was then slowly advanced into cisterna magna through the entire length of the Tuohy Borst adapter and the IV catheter. The adaptor was tightened up, and the baseline ICP referenced to the atmospheric pressure (pre-set to 0 mm Hg) was recorded for 5-15 min. One of the following procedures was subsequently performed.

### ICP setting

A 60" extension set polyvinyl chloride minibore tubing (Acacia) connected to a bottle of sterile saline and subsequently primed was attached to the Tuohy Borst's side arm. ICP registration started immediately after opening the side arm. Initially, the saline surface meniscus was leveled with the cisterna magna thereby equilibrating ICP with the ambient atmospheric pressure. ICPs of 1.0±0.4, 5.5±1.1, 8.3±2.5, 12.3±0.7, 17.6±1.2, 21.1±2.7, 24.7±3.8, 30.1±5.1, and 37.0±2.3 mm Hg were then set in each of the 4 rats by positioning the saline bottle at progressively increasing heights. At each height, ICP was recorded for 10-15 min.

### CSF drainage assessment

A sterile saline bag was suspended on a bench-top pole and attached to a dripper. The dripper was a manually-made device for the flow rate measurement. It consisted of a conical plastic tip hermetically inserted inside the air-filled plastic cylinder. The outside end of the tip was used for the attachment to the saline bag. The opposite end of the cylinder was connected to the saline line leading to the side arm of the Tuohy Borst adapter. When the saline flow was activated by placing a saline bag at different heights, the dripper produced drops of calibrated size (12.8±0.2 ml), the number of which was visually counted using a timer thereby enabling a flow rate calculation. ICP fluctuations caused by falling drops (Fig. 2A) provided an additional method for the dripping rate measurement. Changes to the saline column's heights caused respective changes in the ICP baseline from 0 to 50 mm Hg, which were measured using the ICP registration methodology. For each ICP level, continuous recording was performed for 10-15 min. Experiments were repeated in 4 rats weighing 443±104 g.

### IT infusions with a pump

A Silastic™ silicone tubing (Dow Corning, 0.76 mm inner diameter, 1.65 mm outer diameter), primed and connected to a 1 ml syringe (Becton Dickinson) filled with saline and mounted on a pressure pump (Model 22, Harvard Apparatus), was attached to the Tuohy Borst's side arm. ICP recording was activated immediately with the opening of the side arm and the start of the infusion of 200-240 µl of saline at one of the following rates: 10, 20, 40, 80, and 120 µl/min. Each rate was tested in three different rats. Each of 4 rats was used for 3-4 serial tests allowing sufficient time between infusions as confirmed by the ICP relaxation to the baseline level.

*Manual bolus IT injections*

An infusion plug (Argyle™, Covidien) equipped with a rubber septum was attached to the Tuohy Borst's side arm. A 1-ml tuberculin syringe (Becton Dickinson) filled with saline was inserted in the plug and the injection of a tested volume was performed at 2.7±0.9 ml/min. This is the rate that was typically used in our previous studies utilizing manual cisterna magna injections. The syringe was withdrawn at the end of the injection, and the ICP was allowed to return to the baseline level. The syringe was reloaded, and the injection of the next volume was repeated. The following volumes were tested: 0.05, 0.1, 0.15, 0.2, 0.25, 0.3, 0.35, 0.45,0.6, 0.8, 1.0, and 1.5 ml. For the injection of 1.5 ml of saline, a 3 ml syringe equipped with a 23G needle (Becton Dickinson) was used. A temperature-controlled thermal pad or a heating lamp maintained body temperature over the entire duration of the procedures (ca. 2 h). At the end of each study, a rat was euthanized by way of cardiac infusion of Euthasol (pentobarbital, 100 mg/kg).

## Monkeys

### Sedation and preparation

Animals were fasted for 12 h before the experiment. Atropine sulfate (0.03 mg/kg) was administered IM 1 h before the study to decrease saliva production. 10 min thereafter, the animal was sedated with Ketamine (15 mg/kg)/Xylazine (1 mg/kg) IM at the animal facility and transported to the laboratory. The animal was intubated with a 3 mm ($\varnothing$) endotracheal tube (Mallinckrodt™, Covidien), and the tube was connected to the anesthesia line providing continuous 2% isoflurane/$O_2$ flow at 2 l/min. Heart and respiration rates, blood oxygen saturation ($SpO_2$), end-tidal $CO_2$ content, and rectal temperature were monitored continuously and documented every 15 min. Isoflurane content and flow rate were adjusted upon need to maintain physiological vitals. The anesthetized animal was placed on a soft pad in a prone position. Supplemental heat was provided via a blanket filled with circulating warm water.

### IT injections and ICP measurements

ICP measurements in monkeys were carried out using the equipment configuration shown in Fig. S2. The skin above the IT port was shaved and wiped with 70% isopropanol and treated with Betadine. A saline-primed Huber needle (Access Technologies), attached to the t-connector (Microbore extension set – 5 inch, Hospira) with a clamped catheter, was inserted into the port. The catheter was then attached to the saline-primed side-armed Tuohy Borst adapter (Cook Medical). 1F micro-tip pressure catheter (Millar, Inc) was inserted inside the adapter, down to the Y-junction, and leveled with the surface of the animal's spine. The Tuohy Borst's side arm was then opened to equilibrate the line pressure with the ambient atmospheric pressure. Once a stable pressure line was recorded for at least 10 min, the Tuohy Borst's side arm was closed and the connection with the t-connector was opened enabling indirect in-line ICP ($ICP_L$) recording. Baseline registration was carried out for at least 10-15 min. Continuous

ICP monitoring was then performed during the IT administration of one 5 ml saline bolus in the rhesus monkey and five 1 ml saline boluses, 2.5 min apart or two 3 ml saline boluses, 8 min apart, in the cynomolgus monkey. To emulate the conventionally used procedure, trained personnel performed all IT infusions in monkeys manually. Infusion rates were determined by calculating the ratio of the duration of steady advancement of the syringe plunger to the administered volume. Upon the study completion, anesthesia was gradually withdrawn and the animal was returned to the housing facility.

### Measurement of hydrostatic resistance in the injection port (in vitro)

The same pressure measurement set-up as in the in vivo study with monkeys was utilized. A 23G needle (Becton Dickinson) attached to the 60" extension set polyvinyl chloride minibore tubing (Acacia) was inserted into the t-connector. The other side of the tubing was attached to the 10ml syringe (Becton Dickinson) filled with saline and mounted on a pressure pump (Model 22, Harvard Apparatus). All lumens were primed with saline prior to pressure measurements. Infusion at one of the tested rates was initiated and lasted for $107.8 \pm 37.3$ s until the steady state pressure was established and recorded for $101.0 \pm 37.2$ s. When the infusion was completed, the pressure was allowed to return to the baseline level. $102.0 \pm 42.5$ s after the steady baseline recording, infusion at the next rate was performed. The following infusion rates were tested: 0.01, 0.05, 0.1, 0.2, 0.4, 0.6, 0.8, 1.0, 1.5, 2.0, 2.5, 3.0, 3.5, 4.0, 4.5, 5.0, 5.5, 6.0, 6.5, 7.0, 7.5, and 8.0 ml/min.

### Post-acquisition ICP data processing

Post-acquisition data processing was performed using LabChart 8 software (ADInstruments). Mean ICP was determined by averaging a pressure pulse wave over $4 \pm 1$ min ($1478 \pm 387$ pulses). ICP amplitude is defined as the amplitude of the fundamental (P1) component of the ICP pulse wave. To determine it, fast Fourier transform (FFT) of the ICP time block used for the mean ICP calculation was performed using the following parameters: FFT size: 1000 or 2000 points, data window type: Welch, window overlap: 93.75%. Spectrally resolved P1 harmonic was identified in the frequency range of 3.9-8.8 Hz. ICP relaxation time characterized elasticity of the anatomical components of intrathecal compliance and was measured as a time required for ICP to decrease by 90% from the peak value. Data fitting was performed by a least squares method using a Microsoft Excel 2013 program (linear fitting) or the ZunZun online computational resource ([www.zunzun.com](www.zunzun.com)) (sigmoidal and bi-exponential fitting).

### RESULTS

### Adaptability of rats to the increased ICP

An IT administration normally results in ICP elevations due to the increased volume in the subarachnoid space. Therefore, it was important to investigate how different ICP elevations are tolerated in order to determine safe regimes of IT infusions. To this end, we induced in rats (n=4) progressively increased CSF pressures and monitored the changes in the key ICP parameters: mean ICP, ICP amplitude, and spectral characteristics (frequencies and amplitudes) of the ICP waveform's major components (harmonics).

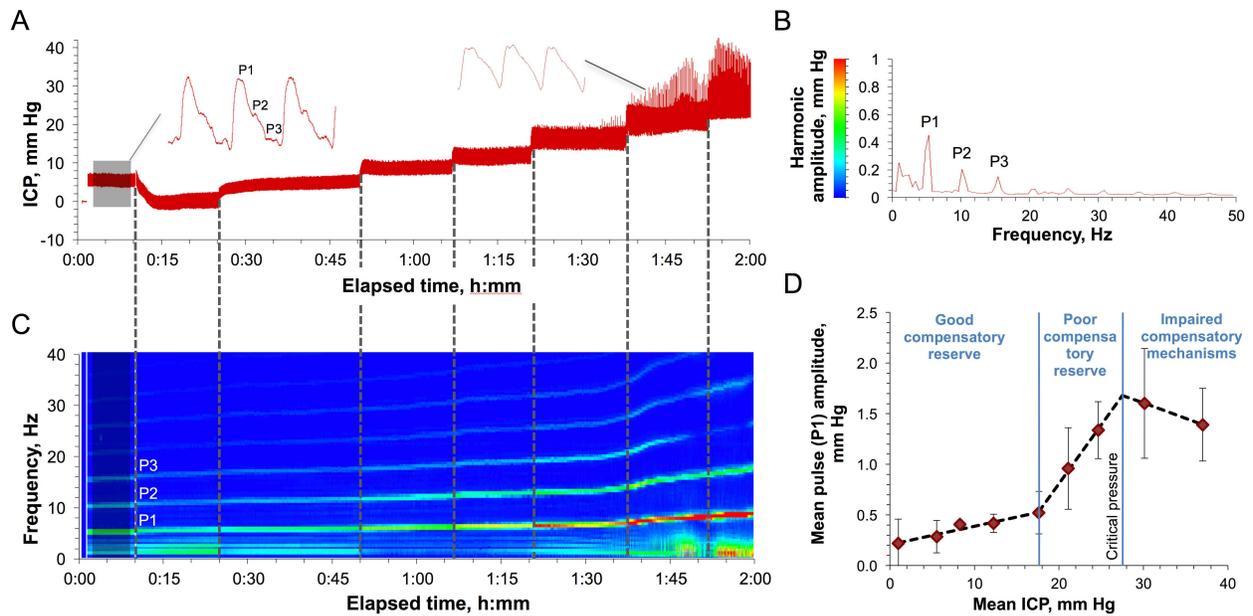

**Figure 1. Characteristics of ICP elevation in a rat. (A)** Various ICPs induced by a saline column attached to the rat's CSF in the cisterna magna. Shaded area indicates baseline ICP. A representative ICP pulse waveform at low and high pressures are shown in the left and right sides, respectively. P1, P2, and P3 are the major pulse's harmonics. **(B)** The amplitude-frequency characteristics of the spectrally resolved P1, P2, and P3 harmonics of the baseline ICP pulses (shaded area in Fig. 1A). A result of averaging of 2564 FFTs each consisting of 2000 points. **(C)** Spectrum view of the changes in the harmonics' frequencies and amplitudes as a result of ICP elevation over time (Fig. 1A). **(D)** ICP autoregulatory capacity and compensatory reserve in rats. Three specific zones characterizing different states of ICP compensatory reserve are shown and are based on the correlation between the mean amplitude of the P1 component of the ICP pulse and the mean ICP. Each data point represents a mean value obtained by averaging over 4 animals. Error bars indicate standard deviations.

Fig. 1 A-C show a representative ICP monitoring experiment in a rat undergoing a progressive ICP elevation. Baseline ICP (Fig. 1A, shaded area) averaging 8.6±1.7 mm Hg (across all studies) was characterized by periodic pulses, a waveform of which had a classical shape (Fig. 1A, left waveform) with spectrally resolved P1, P2, and P3 components (Fig. 1B). ICP elevations caused waveform distortion (Fig. 1A, right waveform) reflecting the changes in the spectral characteristics (frequency and amplitude) of the harmonics as shown in Fig. 1C. Specifically, the harmonics' frequencies and amplitudes, as well as the distance between harmonics increased with ICP elevation, with the most dramatic change affecting the amplitude (5.2±1.5 – fold increase). All the changes became rapidly pronounced after the pressure reached 18 mm Hg, before which only a minor yet consistent increase was observed. Spiking artifacts in the waveforms, which became pronounced at ICP > 20 mm Hg (Fig. 1A), were caused by the increased neck motion of the rats due to the increased respiration rate (visual observations).

The adaptability of rats to the increased ICP (autoregulatory capacity) was characterized by the relationship between the time-averaged amplitude of the P1 component and the mean ICP (Fig. 1D). This relationship has been shown to be a function of the capacity of the compensatory mechanisms to maintain the physiologically relevant pressures in response to different perturbations in the CSF volume[16,17,18]. Correlation between these two variables formed three specific zones characterized by: (1) low synchronization (good compensatory reserve) at low ICPs (1-18 mm Hg), (2) high synchronization (poor compensatory reserve) at higher ICPs (18-28 mm Hg), and (3) negative correlation (impaired compensatory mechanisms) at ICPs > 28±4 mm Hg ("critical pressure"). All rats that experienced ICPs higher than the critical pressure for prolonged time (>10 min) experienced a fatal outcome.

## Performance of the CSF drainage system under pressure

CSF drainage into blood plays a vital role in the accommodation of volumes added to the CSF and in adaptation to elevated ICP due to the pressure-valve function of the drainage sites. The vast majority of these sites are located at the CSF-blood interface in the arachnoid protrusions (arachnoid villi) into the venous sinuses within the cranial (superior sagittal and transverse venous sinuses) and spinal (venous sinuses near dorsal root ganglia) compartments[30,31]. To assess the capacity of these sites under different ICPs, we developed a novel method to measure the CSF drainage rate schematically depicted in Fig. 2A. Briefly, it leverages the measurement of the saline intake in a fluid line connected to the CSF pool, which is challenged by different pressures using a water column effect. Fig. 2B illustrates the obtained ICP dependency of the CSF drainage rate for individual rats (n=4). The dependency has a gross exponential shape with a good linearity in the beginning (6-23 mm Hg). A reciprocal to the linearity coefficient denotes the outflow resistance (0.30 mm Hg/(µl/min)) characterizing the drainage performance. Deviations from linearity in the form of lower ICP than expected at ICP > 23±1 mm Hg were consistent with the respective region of weak compensatory mechanisms (18 mm Hg < ICP > 28 mm Hg) shown in Fig. 1D. Non-linearity can be explained by altered flow characteristics due to either the change in the capacity of the normal CSF drainage sites or the emergence of additional outflow pathways. Maximally tolerated pressures (MTP) marked with black asterisks averaged 30±3 mm Hg correlating with the region of impaired compensatory mechanisms (ICP > 28 mm Hg) shown in Fig. 1D. The following reduction of the drainage rate at ICP > MTP is likely indicative of the discontinuation of some drainage pathways due to death.

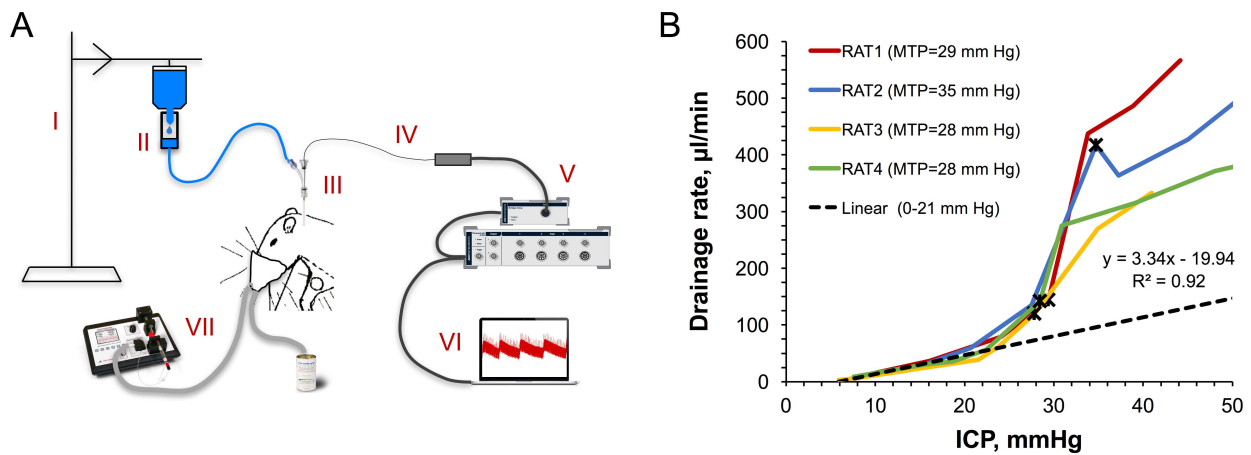

**Figure 2. Viability of the CSF drainage system under pressure. (A)** A sketch of the originally-developed experimental setup for measuring the rate of unconstrained CSF drainage at different ICPs. Components of the setup: (I) bench-top pole with a suspended saline bag set at different heights, (II) dripper with calibrated drop size attached to the saline bag and connected to the saline line, (III) Tuohy Borst adapter attached to the catheter inserted in the cisterna magna; saline line connected to the adapter's side arm, (IV) micro-tip pressure sensor advanced to the cisterna magna through the adapter and catheter, (V) data acquisition platform, (VI) control and data analysis console (note the ICP fluctuations on the computer screen caused by falling drops), (VII) small-animal isoflurane anesthesia platform. **(B)** ICP dependency of the CSF drainage rate. MTP: maximally tolerated pressure. Black

asterisks mark MTP for each rat. Data points in the 6-23 mm Hg region were used to build a linear regression (parameters are shown).

## Pump slow-rate IT infusions in rats

Once the critical pressure, above which adaptation to the increased ICP is impaired, was determined, it was used to determine and characterize safe regimes of IT administrations. To this end, we employed a continuous infusion method when the rats (n=4) were infused 200-240 µl of saline (ca. 100% of the total CSF volume in rats) at different rates while the ICP response was continuously monitored. The obtained data demonstrated that the shape of the ICP buildup is irrelevant of the infusion rate in the tested interval of slow rates (10-120 µl/min) (Fig. 3A). All rates led to the to the ICP stabilization. However, despite different plateau levels, ICP steady state occurred at the same infused volume of 150 µl. The relationship of the plateau ICP with the infusion rate (Fig. 3B) showed a good linearity up to 23.5 mm Hg (40 µl/min), with the linearity coefficient (0.35 mm Hg/(µl/min)) denoting the outflow resistance. It should be noted that both the ranges of the linear response of the CSF drainage system and the values of the CSF outflow resistance obtained by this and the dripping method (Fig. 2B) showed significant correlation thereby validating the latter. Infusion rates higher than 40 µl/min caused significant deviations from linearity in Fig. 3B in the form of lower steady-state ICP than expected. As such, collectively with the data shown in Fig. 2B, 40 µl/min can be considered as the threshold below which infusions are primarily compensated by the fluid outflow through mesothelial pores in arachnoid villi (granulations). The mechanistic basis of the additional drainage system activating at higher rates (and ICP) is to be determined.

Accommodation of liquids added to the CSF at rates > 40 µl/min and ICP autoregulation at pressures > 24 mm Hg appear to be based on a combination of compensatory mechanisms that can be combined in two groups: pressure-dependent CSF drainage and intrathecal compliance. Assessments of their respective involvement can be obtained by analyzing the relaxation of the progressively elevated ICP to the baseline level. For example, 40, 80, and 120 µl/min infusions of 120 µl of saline caused rapid ICP increases above 20 mm Hg followed by the repeated patterns of a multi-phase relaxation process (Fig. 3C). Among two or three relaxation phases, the first and the fastest phase (phase I) had an increasing contribution with the pressure increase. This is likely the result of the action of the mechanisms of intrathecal compliance due to the significantly limited elasticity capacity of the anatomical correlates (not presently clearly established). Analysis of the dependency of the relaxation time on the ICP elevation magnitude (Fig. 3D) obtained for all studies revealed a linear region followed by stabilization at 17.4±0.1 min. The bend point at 12 mm Hg (20.6 mm Hg of the total ICP) corresponding to the infusion rate of 30 mm Hg is likely another indication of the transition from one (mesothelial drainage-based) to two (or more) primary mechanisms involved in the accommodation of liquids added to the CSF at raising infusion rates. In fact, all infusions that caused ICP elevations above 12 mm Hg were followed by multi-phase relaxations.

ICP waveform and frequency analyses of the ICP curve in the representative example given above (Fig. 3C) revealed the presence of P1, P2, and P3 harmonics that remained at their characteristic frequencies (5.4, 10.3, and 15.6 Hz, respectively) (Fig. 3F) during ICP elevation and relaxation phases (Fig. 3E). This observation contrasts with constant-pressure studies (Fig. 1C), which is likely to be explained by the shorter infusion time (1-3 min) seen in the latter case, which was presumably insufficient to increase the heart rate in response to the elevated ICP. The harmonics' amplitudes, in contrast, promptly increased with each ICP elevation (Fig. 3E, color transitions from green to red) suggesting the robust impact of these infusions on the compliance-based compensatory reserve, which agrees with our findings shown in Fig. 1D (poor compensatory reserve region).

Time-dependency of the response of different physiological components to elevated ICP turned out to play a critical role in the safety of high-volume intrathecal boluses as follows below.

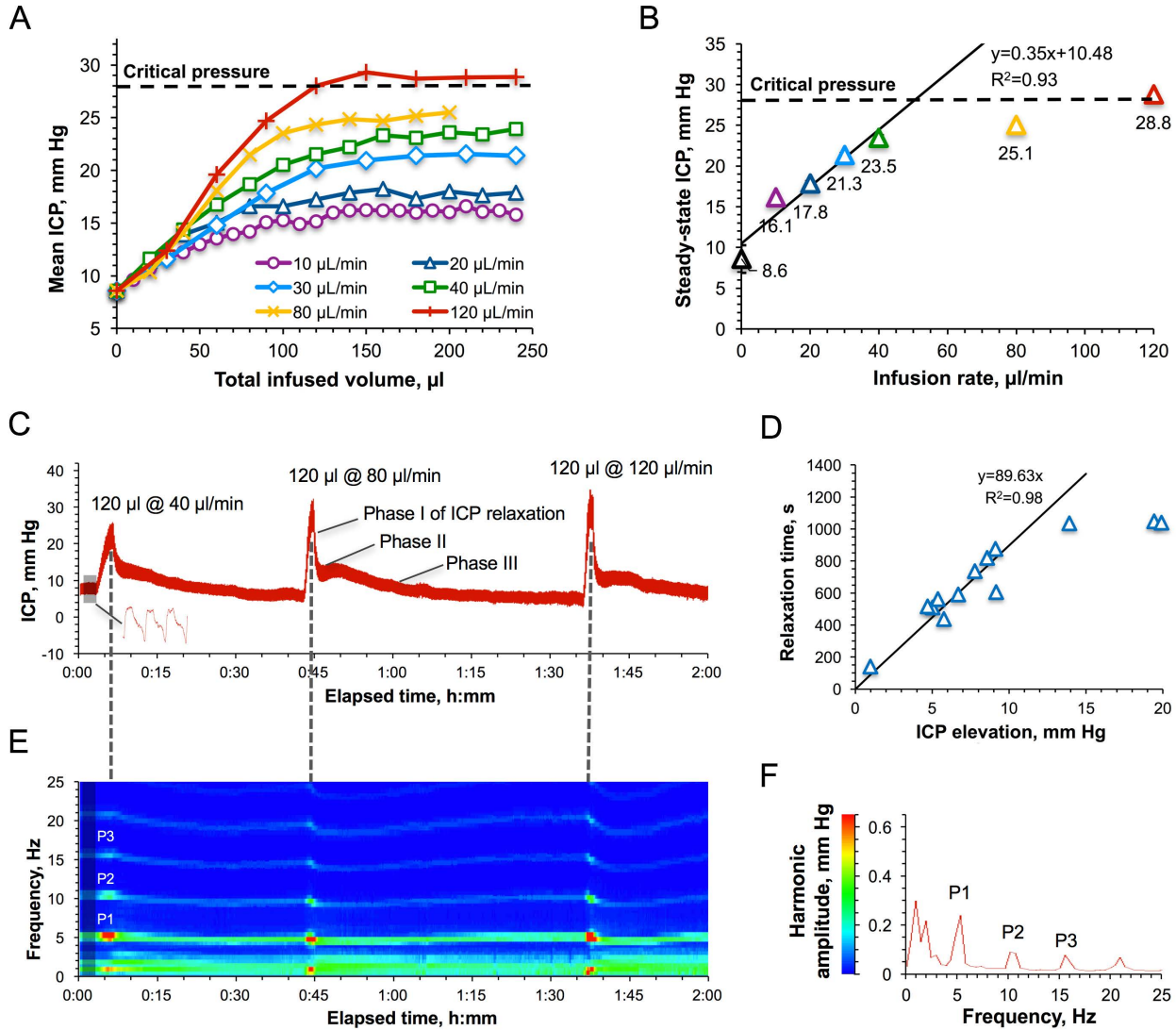

**Figure 3. ICP response to IT infusions. (A)** ICP – IT infused volume relationship for different infusion rates. "Critical pressure" (28±4 mmHg) denotes the ICP above which pressure compensatory mechanisms were found to be impaired (Fig. 1D). **(B)** Relation of the approximated plateau ICPs with the infusion rates. Parameters of the linear regression for infusion rates in the interval of 10 – 40 µl/min are shown. Error bars represent standard errors obtained by averaging the data points in Fig. 1A at the steady state (> 150 µl of the infused volume). **(C)** A representative picture of the ICP response to the infusion of the same volume (120 µl) at three different rates - 40, 80, and 120 µl/min. The shaded area is the baseline ICP. A representative ICP waveform is shown in the bottom. **(D)** Relationship between the time of relaxation to the baseline ICP and the magnitude of ICP elevation. Non-averaged data points of individual experiments are shown. Parameters of the linear regression are given for 10 – 40 µl/min infusion rates. **(E)** A representative picture of the harmonic analysis of the shaded area in Fig. 3C. A result of averaging of 1158 FFTs each consisting of 2000 points. P1, P2, and P3 harmonics were spectrally resolved at all pressures. **(F)** Spectrum view of changes in the harmonics' frequencies and amplitudes as a result of IT infusions.

## Manual bolus IT injections in rats

Sequential injections of saline volumes ranging from 0.05 to 1.5 ml at the average rate of 2.7±0.9 ml/min in the cisterna magna of a rat resulted in the prompt ICP elevations (Fig. 4A). The dependence of the magnitude of those elevations on the injected volume had a logarithmic shape (Fig. 4B). ICP stabilized at ca. 200 mm Hg level, which was found to be the maximally-tolerable suggesting the capacity limit of the relevant adaptation mechanisms. ICP relaxations to the baseline level (Fig. S3) were of bi-exponential shape as evidenced by moving frame linearizations (Guggenheim method[32]) having a bi-linear shape (Fig. S4). Bi-exponentiality reflects the involvement of fast and slow relaxing components characterized by large and small time constants. Relative fractions of both components as a function of the ICP elevation magnitude are shown in Fig. S9A revealing the diminished contribution of the slow relaxation processes as the ICP increased. It is of note that the relationship of the total relaxation time (ranging from 9 to 172 s) with the ICP elevation magnitude had a bi-exponential form whereas the same relationship of the total elevation time (2-20 s) was linear (Fig. 4C). Neither ICP elevations nor relaxations were associated with significant changes to the heart or respiratory rates (Table S1).

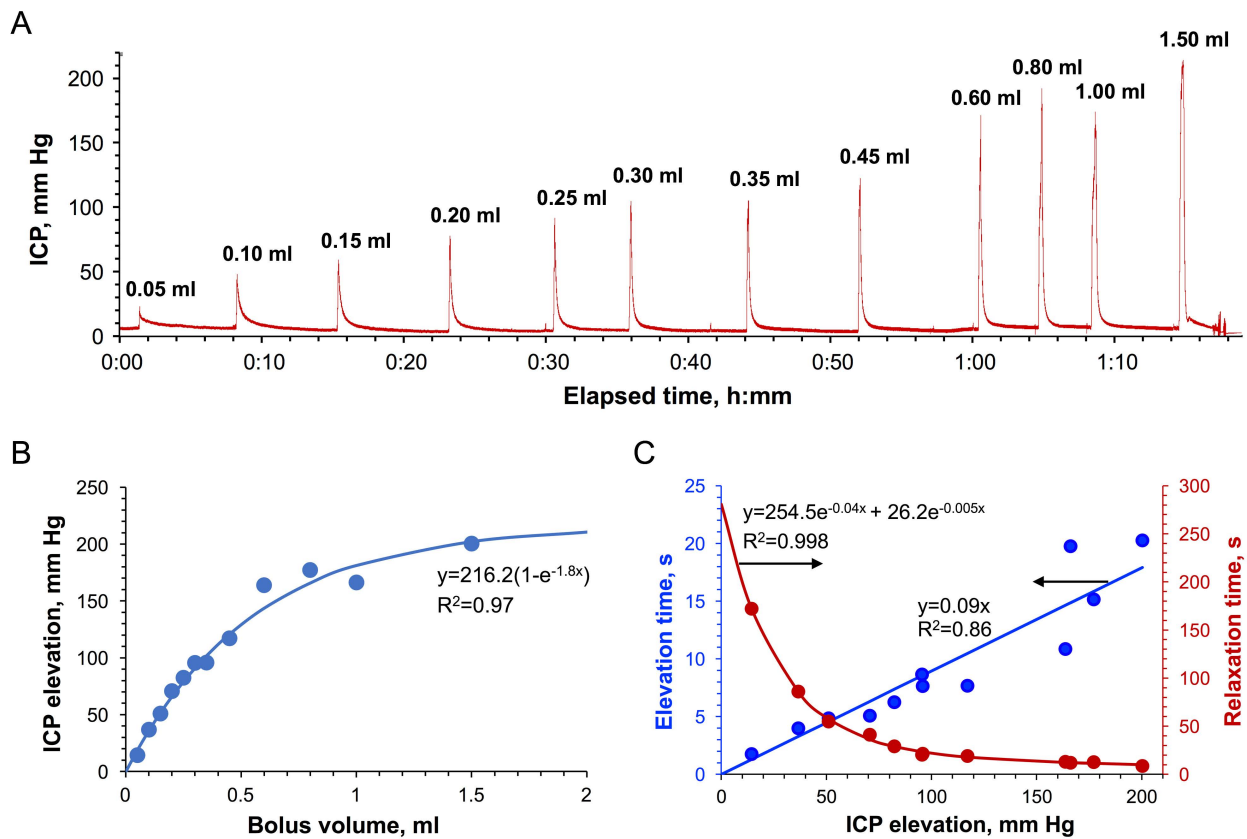

**Figure 4. Bolus IT injections in rats. (A)** The time-dependence curve of ICP elevation. The average administration rate is 2.7±0.9 ml/min. **(B)** Dependence of the ICP elevation on the injected volume. **(C)** The pressure dependence of the total ICP elevation (blue) and ICP relaxation time (red).

## Large-volume manual IT administrations in ported monkeys

From the translational perspective it was important to ascertain if the ICP effects observed in rats are similar to those in monkeys. The impact of large-volume (with respect to the total CSF volume) IT infusions on the ICP in monkeys was assessed by monitoring the $ICP_L$ in the saline-filled catheter attached to the IT port during and after manual bolus administrations of 1, 3, and 5 ml (ca. 5-25% of the CSF volume) of saline (Fig. 5, A-C). All injections caused the initial $ICP_L$ elevation from the baseline level of $2.2\pm0.7$ mm Hg to 85-200 mm Hg. Once each injection was completed and a needle was withdrawn, $ICP_L$ quickly decreased to 30-60 mm Hg followed by a subsequent significantly slower relaxation to the initial baseline $ICP_L$ accompanied by pressure pulsations. Such a complex $ICP_L$'s behavior can be explained by the fact that $ICP_L$ is made up by two components – the endogenous CSF pressure and hydrodynamic resistance pressure in the injection system. The latter component can be deteriorated by the partial catheter compression due to the tissue build-up around the catheter over time. However, the bottom estimate of the dependence of the resistance pressure on the infusion rate was obtained in the in vitro study using the same port and $ICP_L$ measurement method (Fig. 5D). The best linear fit was found for the 0-7 ml/min interval of injection rates. Time dependence of the $ICP_L$ during infusion (Fig. S5, A) at each tested rate from 0.01 to 8 ml/min demonstrates quick (2-13s) establishment of the steady-state pressures characterizing hydrodynamic resistance. Relaxation to the baseline ICP after the infusion completion was somewhat slower and ranged from 8s to 21s. Both elevation and relaxation durations were found to be a linear function of the infusion flow rate (Fig. S5, B).

Characterization of the infusion resistance pressure in the port allowed discrimination of the real pressure in the CSF. ICP dependence on the infused volume for the studied infusion rates is shown in Fig. 5E. Infusions with rates up to 2 ml/min revealed the tendency to lead to steady-state pressures, which were found to linearly depend on the infusion rates (Fig. 5F), as in the rat studies. Over the entire duration of the procedures, the monkey's vital parameters (heart and respiration rate, $SpO_2$, $CO_2$ content in the exhaled air, and rectal temperature) were constantly monitored. Respiration and cardiac complications (elevations of both rates followed by apnea) became apparent at pressures exceeding 50 mm Hg. Due to this safety limit, we were not able to formally characterize administrations of volumes higher than 1 mL for high-rate infusions (3.4-7.8 ml/min, Fig. 5A and E).

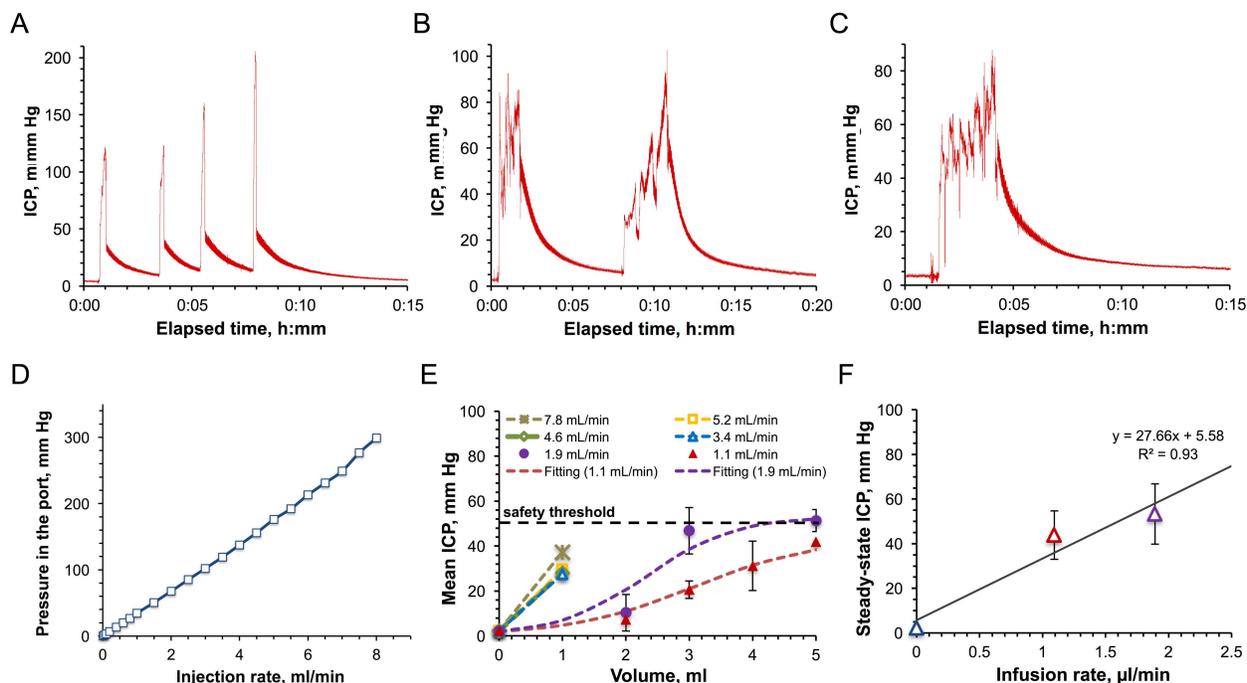

**Figure 5. Large-volume bolus IT injections in monkeys.** Time-dependence of the ICP measured in-line (ICP$_L$) for a series of 1 ml bolus injections in the cynomolgus monkey with the average infusion rate of 5.3±1.8 ml/min **(A)**, a series of 3 ml infusions in the cynomolgus monkey with the average infusion rate of 1.9±0.6 ml/min **(B)**, and a 5 ml bolus injection in the rhesus monkey with the infusion rate of 1.8 ml/min **(C)**. **(D)** Relation of the IT injection port's ICP$_L$ with the injection rate. **(E)** Dependence of the CSF pressure (ICP$_{CSF}$) on the infused volume for different infusion rates. The ICP$_{CSF}$ was derived by subtracting the resistance pressure in the port. **(F)** Relation of the approximated steady-state ICPs with the infusion rates. Parameters of the linear regression are shown. Error bars in (E) and (F) represent standard deviations.

As in the rat studies, the pressure relaxation kinetics in NHPs was found to be best fit by a bi-exponential curve (Fig. S6) as suggested by the bi-linear shape of the Guggenheim linearization (Fig. S7). Bi-exponentiality reflects the involvement of fast and slow relaxing components characterized by large and small time constants. Relative fractions of both components as a function of the ICP elevation magnitude are shown in Fig. S9B revealing the diminished contribution of the slow relaxation processes as the ICP increased. The total time of ICP relaxation from the peak value to the baseline level averaged 5.0±0.6 min without a noticeable correlation with the ICP elevation magnitude.

## DISCUSSION

This study was aimed at estimating the safety thresholds for IT infusions in rats and monkeys based on ICP-derived indices. In the first set of studies in rats, the animals were subjected to gradually increasing pressures applied within the CSF while measuring major ICP waveform characteristics, such as a mean ICP and a pulse amplitude of the fundamental component (P1, percussion wave). Analysis of the relationship between these two characteristics is frequently used for the assessment of the ICP autoregulatory capacity and the compensatory reserve[16,17,33]. These concepts characterize the physiological mechanisms of adaptation to the increased ICP. Exhaust of these resources is typically associated with irreversible changes due to the high pressures and leads to poor outcomes[16,17].

Published analytical[34] and experimental[19] data demonstrate a linear relationship between the ICP pulse amplitude and the mean ICP. Additionally, detailed analysis of the sigmoid form of the ICP-CSF volume curve suggests three different degrees of linearity specific for three different ICP zones[17,19,33]. The first zone, situated at low pressures, is characterized by low synchronization between changes in amplitude and a mean ICP, indicating a good compensatory reserve used to accommodate volume alterations occurring as a result of cerebral blood pulsations. In the middle zone, this synchronization is more pronounced and therefore the pulse amplitude increases more readily with the raised ICP. This denotes a poor compensatory reserve, implying that even small volume augmentations can become uncompensated trending toward rapid ICP elevations that can become dangerous in nature. Negative correlation at high ICPs denotes detrimental, irreversible changes in the compensatory mechanisms leading to permanent structural damages or death.

Our rat data (Fig. 1D) is consistent with the described amplitude-pressure paradigm showing a relatively broad (18 mm Hg) area of low synchronization and similar middle and third zones (ca. 10 mm Hg-wide each). By combining the first and second zones, one can derive the working range of safe ICPs for performing IT administrations. This relatively wide range spans from the baseline level of $8.6 \pm 1.7$ mm Hg through the critical ICP of $27.5 \pm 3.8$ mm Hg. Animals that underwent pressures higher than the critical ICP for prolonged periods of time (10-15 min) experienced a fatal outcome likely associated with cardiac complications. In fact, during ICP elevation we observed a progressive shift of the fundamental (P1) component of the ICP pulse wave to higher frequencies (from 5.4 Hz to 8.3 Hz) (Fig. 1C) indicating an increased heart rate (Fig. S8).

The baseline and "critical" ICP values obtained in this study are also in good agreement with the published data, which specify a relatively wide range of pressures with respect to species, measurement method, posture, and health condition. For example, the values of the resting CSF pressure reported for the prone position span from 3 to 8 mm Hg in anesthetized adult rats (cisterna magna measurements)[35.36,37], from 7 mm Hg to 14 mm Hg in non-anesthetized monkeys (cisterna magna measurements)[38], and in non-anesthetized humans from 7 mm Hg to 15 mm Hg (lumbar measurements)[20]. As a criterion of tolerance for elevated ICPs, it was proposed[39] to use a linearity of the plateau ICP-infusion rate relationship derived from a constant-rate infusion experiment. It was suggested[39] that deviations from linearity at high rates, in the form of lower plateau ICP than expected (decrease in the CSF outflow resistance) could be indicative of altered flow characteristics of the CSF outflow sites. Most of these sites are located on the CSF-blood interface in the arachnoid protrusions (arachnoid villi) within the venous sinuses in the cranial (superior sagittal and transverse venous sinuses) and spinal (venous sinuses near dorsal root ganglia) compartments[30,31]. At high flow rates, unidirectional and CSF pressure-dependent[31] flow through arachnoid villi can reach intrinsic capacity thereby hypothetically increasing the risk of structural damages to the drainage components. For instance, ultrastructural changes in the arachnoid villi of adult rats have been reported for CSF pressures above 22 mm Hg[40]. A linear plateau pressure response to constant-rate infusions with maximal ICP ranges from 15 to 30 mm Hg, has been found for a number of adult animals (mice[41], rats[39], cats[42], rabbits[43], monkeys[44]) and man[20,45,46]. A non-linear response has even been observed in rats at low infusion rates (<34 μl/min) although the plateau ICP ranged from 29 mm Hg to 81 mm Hg in that study[35]. At rates higher than 34 μl/min, equilibrium pressures were not consistently obtained, and electroencephalographic patterns began to reflect the disturbed cerebral function. Therefore, acquisition of steady-state pressures became a widely-accepted criterion of physiologically safe adaptation to the slowly infused liquids regardless of the shape of ICP-infusion rate relationship. No such criterion was documented for bolus injections however.

Considering that a constant-rate infusion method is associated with forced perfusions through the LMS, which could hypothetically be damaging per se at high rates, we developed the alternative method, which harnessed a direct measurement of the rate of unconstrained, free drainage at varying ICP values (Fig. 2). The obtained gross exponential form of the drainage rate-ICP relationship suggests a high throughput of the CSF drainage system, which hypothetically consists of several contributary factors activating at different pressures. A good linear regression obtained until the ICP of 23±1 mm Hg can be attributed to drainage through the arachnoid villi based on the previously reported data[39] and correlation with our findings in the constant-rate infusion study (Fig. 3B). However, further deviations from linearity occurring from 23±1 mm Hg to 30±3 mm Hg (MTP) in Fig. 2B and similarly from 23.5 mm Hg to 28.8 mm Hg (critical pressure) in Fig. 3B are likely explained by the emergence of additional, extra-villious drainage pathways rather than previously suggested pressure-related ultrastructural damages to arachnoid villi[40]. In fact, deviations from linearity were still characterized by the steady state ICP condition at the respective pressures (Fig. 3A), which is suggestive of physiologically normal functionality of the underlying mechanisms. Furthermore, the ICP relaxation process repeatedly returned the elevated ICP to its baseline value (Fig. 3C), albeit at faster rates (Fig. 3D), which would not have been the case if any damages had occured. Additionally, it should be noted that the entire interval of tolerated pressures (6-30 mm Hg) in Fig. 2B correlates with the range of unimpaired compensatory mechanisms (9-28 mm Hg) in the constant pressure study (Fig. 1D). Lethality at pressures averaging 30±3 mm Hg occurred only after prolonged (>10 min) exposures and is likely explained by extended, non-elastic (as suggested by negative correlation in the respective ICP region on Fig. 1D) compressions of vital CNS or vascular components (requires further investigation). The following reduction of the drainage rate at ICP > MTP is likely indicative of the discontinuation of some drainage pathways due to death.

The exact mechanisms of the CSF drainage through the mesothelial layer of arachnoid villi are insufficiently elucidated due to significant methodological difficulties. The working hypotheses include an active vesicular transport[47] such as a fluid-phase caveolae-mediated macropinocytosis[48,49] and filtration through either permanent or dynamic mesothelial pores formed by merged vacuoles[50]. A recent study of the pressure relief mechanisms in the endolymphatic sac of the inner ear[51] adds the theoretical possibility of a similar passively working pressure valve in the mesothelial layer of arachnoid villi. In the inner ear, the valve is formed by the dynamic overlapping basal lamellae of adjacent endothelial cells, the apical junctions of which are separated under increasing pressure thereby forming ~1 μm-wide gaps. The authors specifically note similarities between the ultrastructure of their lamellar barriers and the previously described giant vacuoles found abundantly in the pressure- and fluid traffic-regulating tissues, such as Schlemm's canal of the eye[52], endolymphatic sac's epithelium of the inner ear[53], and the mesothelial layer of arachnoid granulations[50]. Identification of the LMX1B transcription factor involved in the apical junction remodeling, a mutation of which causes various symptoms (e.g., glaucoma[54], nephropathy[55], and hearing impairment[55]) associated with impaired fluid transport and hypertension in various tissues, is also suggestive of the common underlying pressure regulation mechanisms. Fig. 2B might also suggest that active and passive drainage mechanisms work in concert in the living body. In fact, average post-mortem reduction of the unconstrained drainage can be explained by the dysfunction of active processes and the remaining functionality of the drainage components that are based solely on physical principles (permanent pores). Research into the relevance and relative contributions of the proposed mechanisms remains in high demand.

The role and the location of the extra-villous CSF absorption pathways remain poorly investigated. The most widely discussed route is the hypothesized connection between the CSF and the lymphatic system. CSF absorbed by the olfactory mucosa and cranial and spinal nerve sheaths is believed to be further drained by the lymphatic system.[56] The amount of the CSF drained through this pathway is highly debatable, varying from negligible or little[57,58] to quite significant (up to 50%)[59,60], and appears to be species-dependent[58]. Whatever the fraction of CSF is drained at normal CSF pressure, elevated ICP can potentially increase it[61]. For instance, a study in sheep demonstrated that only 10% of cervical lymph was derived from the CSF at normal ICP, but as ICP increased 7-fold, 80% of cervical lymph was derived from CSF and the cervical flow rate increased 4-fold.[62] The functional role of the lymphatic pathways in man remains poorly investigated, however, primate data[58] suggests negligible contributions. As such, attention should be paid when extrapolating the pressure data between species.

Overall, the obtained data is consistent with previous reports and suggests the primary role of CSF drainage mechanisms in adaptation to raised ICP caused by low-rate infusions. The resulting equilibration of the infusion rate and the pressure-dependent drainage rate prevents liquid and pressure buildup and enables continuous perfusion within the LMS. We found that in rats all infusions leading to the plateau pressures could safely be used for IT administration of at least 240 µl (ca. 100% of the CSF volume in rat[35]). Although the upper volume limit remains to be determined, it is likely to depend more on the compound's cytotoxicity and hemoconcentration rather than volumetric limits of the LMS. This observation lays the groundwork for the development of the CSF dialysis technology, which has the potential to become a novel treatment method as noted recently[63].

High-rate infusions and bolus injections (2.7±0.9 ml/min in rats and 5.3±1.8 ml/min in monkeys) caused rapid uncompensated fluid and pressure buildup, which caused different physiological reactions in rats and NHPs. The upper ICP threshold for bolus injections in rats was 200 mm Hg, which corresponded to a bolus volume of 1.5 ml (Fig. 4B). Short exposure time appeared to play a critical role in the tolerance of these pressures in rats, as no respiratory or heart rate changes were observed (Table S1). In monkeys, the upper ICP threshold was limited to 50 mm Hg regardless of the exposure time and was related to immediate respiratory and cardiac complications (elevations of both rates followed by apnea) demanding resuscitation interventions. The cardiac complications can be somewhat attributed to the partial compression of cerebral vessels. Our recent study[64] showed that in monkeys, the vascular factor contributed ca. 10% to compliance at pressures higher than the diastolic pressure (30-50 mm Hg under isoflurane anesthesia).

Acute uncompensated ICP elevations suggest the minimal contribution of the drainage system and the primary role of the components of craniospinal compliance in adaptation to bolus administrations. These components involve mechanical entities capable of limited expansion and contraction. These physical factors of adaption contribute to the fast ICP elevation (<20 s) and relaxation (<200 s) times (Fig. 4C) suggesting low elasticity capacity of the involved components. This contrasts with physiological factors (drainage systems), for which significant activation times (1-15 min) were observed in rats (Fig. 3A). Although the exact locations of the entities contributing to the hydrostatic compliance have not yet been established, these hypothetically include membranes of the atlanto-occipital and atlanto-axial joint capsules, membranes covering openings (foramina), through which nerves and blood vessels exit the LMS, cerebral and spinal nerve sheaths, and fluid compartments of the inner ear. Whereas enlargement of the optic nerve's meningeal sheath at elevated ICP is well documented[65,66], the behavior of other nerves' sheaths is yet to be established. In rodents, the nerve bundles exiting the cribriform plate are significantly more developed than those in primates and therefore their contributions to volume and pressure attenuations may be expected at a higher degree. The role of the inner ear is mediated by the established communication via the cochlear aqueduct between the subarachnoid space of the posterior cranial fossa and the perilymphatic space of the cochlea. Widely open in most mammals and variably patent in humans (89% of young adults and 70% of older adults[67]), the cochlear aqueduct is situated in the petrous part of the temporal bone and therefore is incapable of any expansions. This rigid structure facilitates instant pressure transmission from the CSF to the fluids of the inner ear.[68] However, compliant cochlear membranes (oval and round windows, tympanic, Reissner's and basilar membranes) can plausibly accommodate some minor volume and pressure changes. Such compliance is believed to be critical for reducing the risk of structural damages to the cochlea due to relatively large and rapid ICP changes as a result of everyday events such as coughing and sneezing.[69] These functions of intra-cochlear hydromechanics underlie the novel emerging method of non-invasive ICP monitoring based on the measurements of the tympanic membrane displacement.[70]

How a system returns to the initial state after perturbations is an important parameter of its adaptive capacity and can also be indicative of the involved mechanisms. In this respect, ICP relaxation to the baseline level after low- and high-rate infusions was analyzed in rats and monkeys. Low-rate infusions characterized by the steady-state ICP demonstrated the longest pressure relaxation durations (141-1047 s in rats and 300±36 s in monkeys). This observation is consistent with the primary contribution of the slow-responsive CSF drainage system to the adaptation of this mode of administration. However, this contribution was gradually diminished with the increase of the infusion rate as indicated by the distortion of the shape of the relaxation curve by the appearance of the quick-relaxation phase (Phase I on Fig. 3C). Analysis of the purely bi-exponential relaxation curves specific for bolus injections in rats (Fig. S3 and Fig. S4) and monkeys (Fig. S6 and Fig. S7) clearly revealed the ever increasing role of the rapidly relaxing component with the raise in the ICP elevation magnitude (Fig. S9). This component can likely be attributed to the mechanisms of craniospinal compliance, very limited elasticity capacity of which explains the fast ICP elevation (2-20 s) and relaxation (9-172 s) times after bolus injections in rats. The gradual transition from the leading role of CSF drainage to compliance mechanisms was also illustrated by the change of the respective relationships between the relaxation time and the ICP elevation magnitude. Whereas it is linear at low-rate infusions (Fig. 3D), it is reciprocally bi-exponential at high rates (Fig. 4C), with the transition occurring within the 15-20 mm Hg region.

Comparison of the volume and pressure tolerance characteristics in rats and monkeys reveals many similarities though, it is essential the differences be emphasized. The same shape of the pressure-volume curves at low infusion rates suggests common underlying CSF drainage mechanisms. However, in rats these mechanisms could possibly be supplemented with lymphatic (or other) drainage pathways. Increased cardiac and respiratory rates at pressures approaching, or exceeding, the safety thresholds suggest the involvement of vascular reactivity, which in rats was significantly slower as evidenced by the time dependence of adversity. However, the most dramatic differences were observed in the tolerance of pressures exceeding the safety thresholds, which was specific for large-volume bolus injections. These pressures in monkeys (>50 mm Hg) immediately caused apnea, likely explained by the excessive pressure on the respiratory center in the medulla oblongata and pons. This effect in rats was not observed until the ICP of 200 mm Hg, which suggests fundamental physiological differences critical for translational studies.

## CONCLUSIONS

ICP monitoring is a valid method for determining safety thresholds for intrathecal therapy delivery. Craniospinal compliance and CSF drainage to the blood are the primary mechanisms of adaptation to elevated CSF pressures. However, their relative contributions, capacity, and physiological correlates are species-dependent and therefore caution should be used in extrapolation of the results between species. It is clear that volume and infusion rates define the duration and magnitude of CSF pressure elevation and both factors must be considered in the optimization of delivery strategies with respect to safety thresholds. While very limited elasticity capacity of the components of craniospinal compliance facilitates solute distribution in the CSF, it is important to pay attention to not exceed the safety thresholds at high-rate administrations. Notably, a pathological process or trauma may impact any of those thresholds and therefore the disease's potential influence should be considered each time IT therapy delivery is to be used. Availability of methodologies enabling fast and reliable assessment of safety thresholds would significantly facilitate both preclinical research and, in a clinical setting, development of delivery strategies optimized for individual patients from both efficacy and safety perspectives.

## AUTHOR CONTRIBUTIONS

VB conceived, designed and performed all the rat studies, performed ICP measurements in NHPs, analyzed and interpreted all ICP data, and drafted the manuscript. JA assisted with the development and carrying out of ICP measurements in rats, performed NHP studies, and analyzed the ICP data. SL performed ICP measurements in rats and analyzed the data. PG and BA assisted with ICP procedures and data analysis. MP conceived and designed NHP studies, assisted with procedures, and interpreted the data.

## DATA AVAILABILITY

The raw data supporting the conclusions of this manuscript will be made available by the authors, without undue reservation, to any qualified researcher.

## CONFLICT OF INTEREST

All authors declare no competing interests.

## FUNDING

This study was supported by the MGH ECOR Formulaic Award (PI: Belov) and the NIH grants R01NS092838 (PI: Papisov) and R21NS090049 (PI: Papisov).

## ACKNOWLEDGEMENTS

Authors thank Fredella Lee for assistance with animal experiments.

## SUPPLEMENTAL MATERIALS

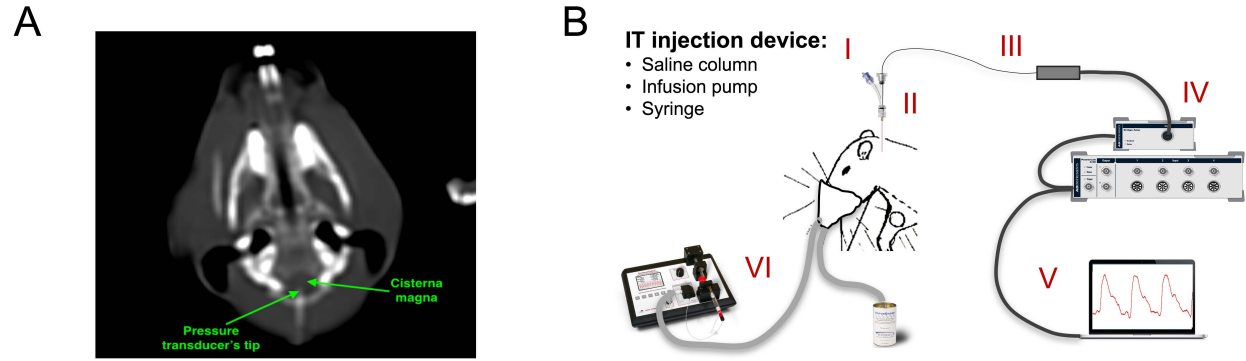

**Figure S1. ICP measurements in rats. (A)** Post-mortem CT scan of the atlanto-occipital junction verifying the catheter positioning in the cisterna magna. **(B)** A sketch of the equipment configuration, which includes the following components: (I) IT injection device (saline bag set at different heights, infusion pump, or syringe for manual injections), (II) Tuohy Borst adapter attached to the catheter inserted into the cisterna magna, with injection device connected to the adapter's side arm, (III) micro-tip pressure sensor advanced into the cisterna magna through the adapter and catheter, (IV) data acquisition platform, (V) control and data analysis console, (VI) small-animal isoflurane anesthesia platform.

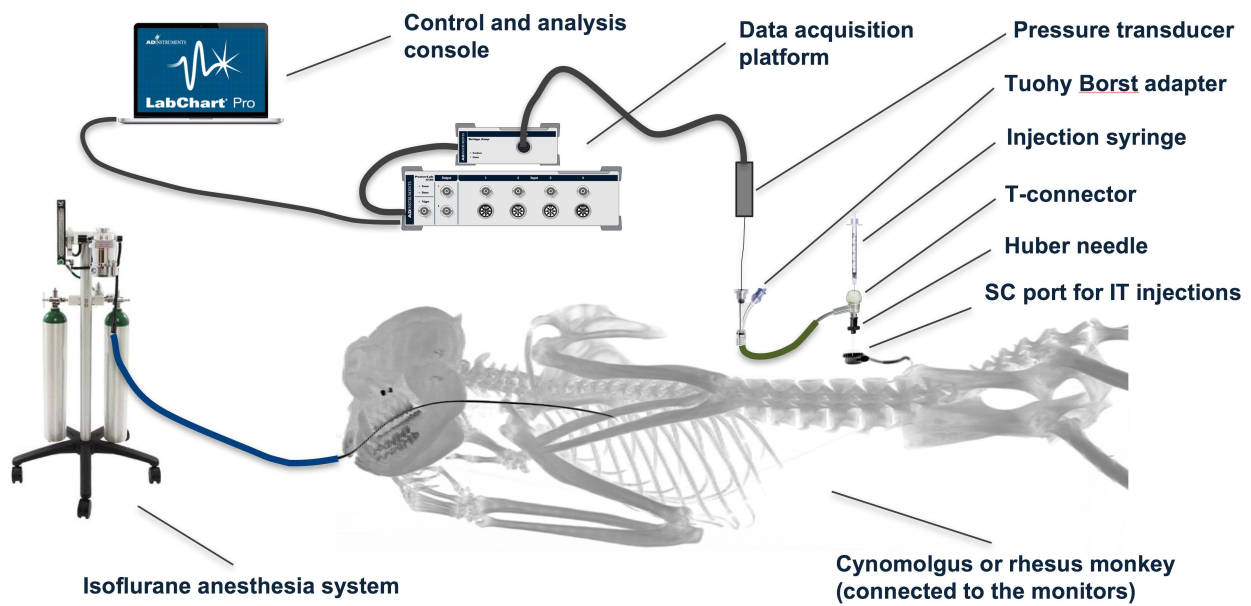

**Figure S2. ICP measurements in monkeys.** A sketch of the equipment configuration.

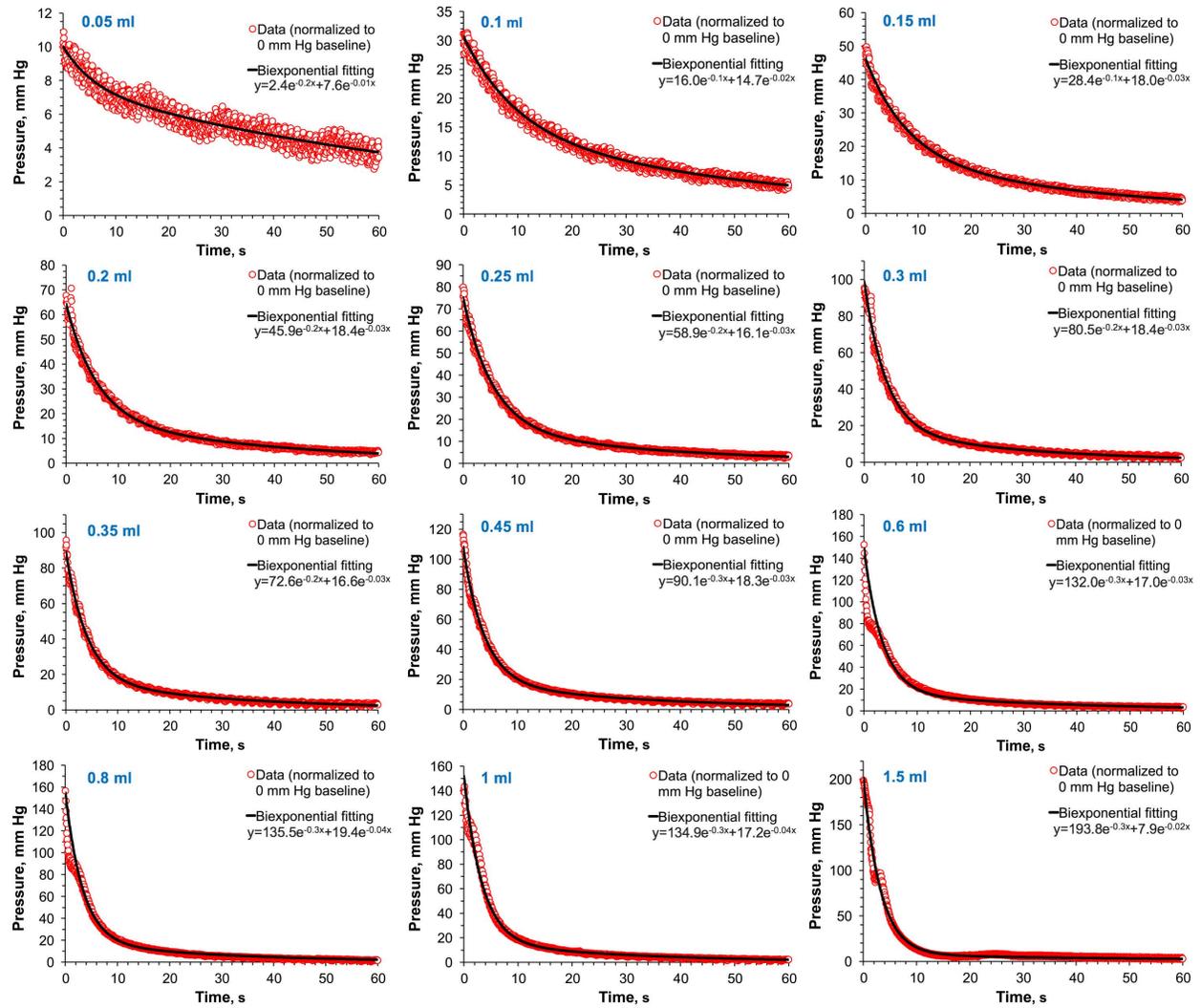

**Figure S3. ICP relaxation to the baseline value after bolus IT injections in the cisterna magna of a rat.** Left to right: time-dependence curves of ICP (blue dots) and their bi-exponential fitting (red lines) after reaching the peak ICP as a result of administrations at 2.7±0.9 ml/min of 0.05 to 1.5 ml of saline. A time interval of 60s is shown. All data are normalized to 0 mm Hg baseline to account for differences in resting pressures.

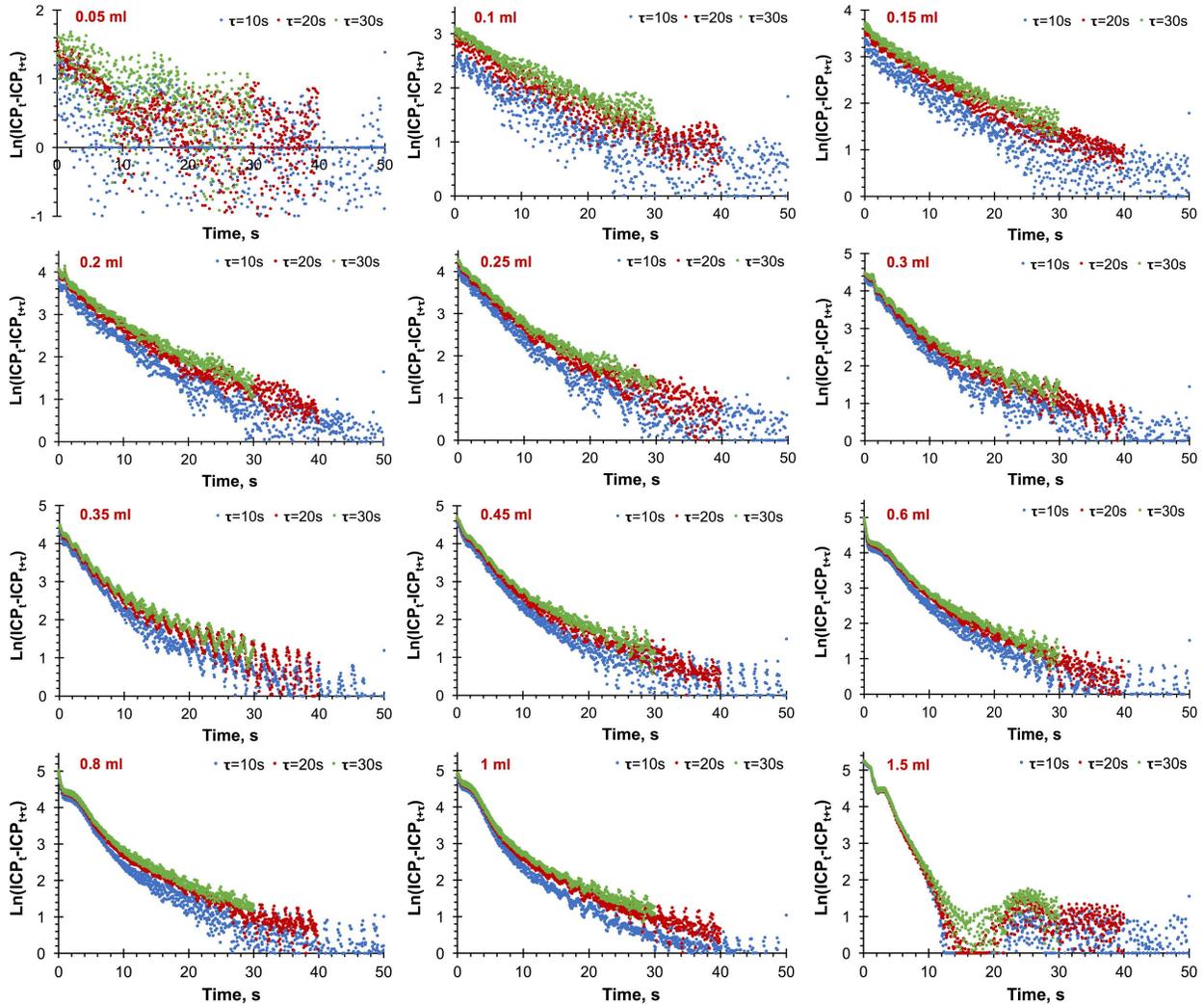

**Figure S4. Guggenheim linearization of the ICP relaxation after bolus IT injections in rats.** Left to right: time dependence of the logarithm of difference between two ICP values on the ICP relaxation curves (Fig. S1), which are 10s (blue dots), 20s (red dots), and 30s (orange dots) apart, after administrations at 2.7±0.9 ml/min of 0.05 to 1.5 ml of saline. A time interval of 50s is shown. Bi-linear shape is indicative of bi-exponential character of the ICP relaxation.

| Volume, ml | Elevation | | Relaxation | |
|---|---|---|---|---|
| | Heart rate, bpm | Respiration rate, bpm | Heart rate, bpm | Respiration rate, bpm |
| 0.05 | 293 | 60 | 293 | 60 |
| 0.10 | 234 | 60 | 234 | 60 |
| 0.15 | 234 | 60 | 234 | 60 |
| 0.20 | 234 | 60 | 234 | 60 |
| 0.25 | 234 | 60 | 234 | 60 |
| 0.30 | 234 | 60 | 294 | 60 |
| 0.35 | 294 | 60 | 354 | 60 |
| 0.45 | 354 | 60 | 354 | 60 |
| 0.60 | 354 | 60 | 354 | 60 |
| 0.80 | 354 | 60 | 354 | 60 |
| 1,00 | 354 | 60 | 354 | 60 |
| 1.50 | 354 | 60 | 294 | 60 |

**Table S1. The effect of bolus IT injections on the heart and respiration rates in rats.** Heart and respiration rates in beats per minute (bpm) are provided for different injected volumes (0.05-1.5 ml) during ICP elevation and relaxation phases.

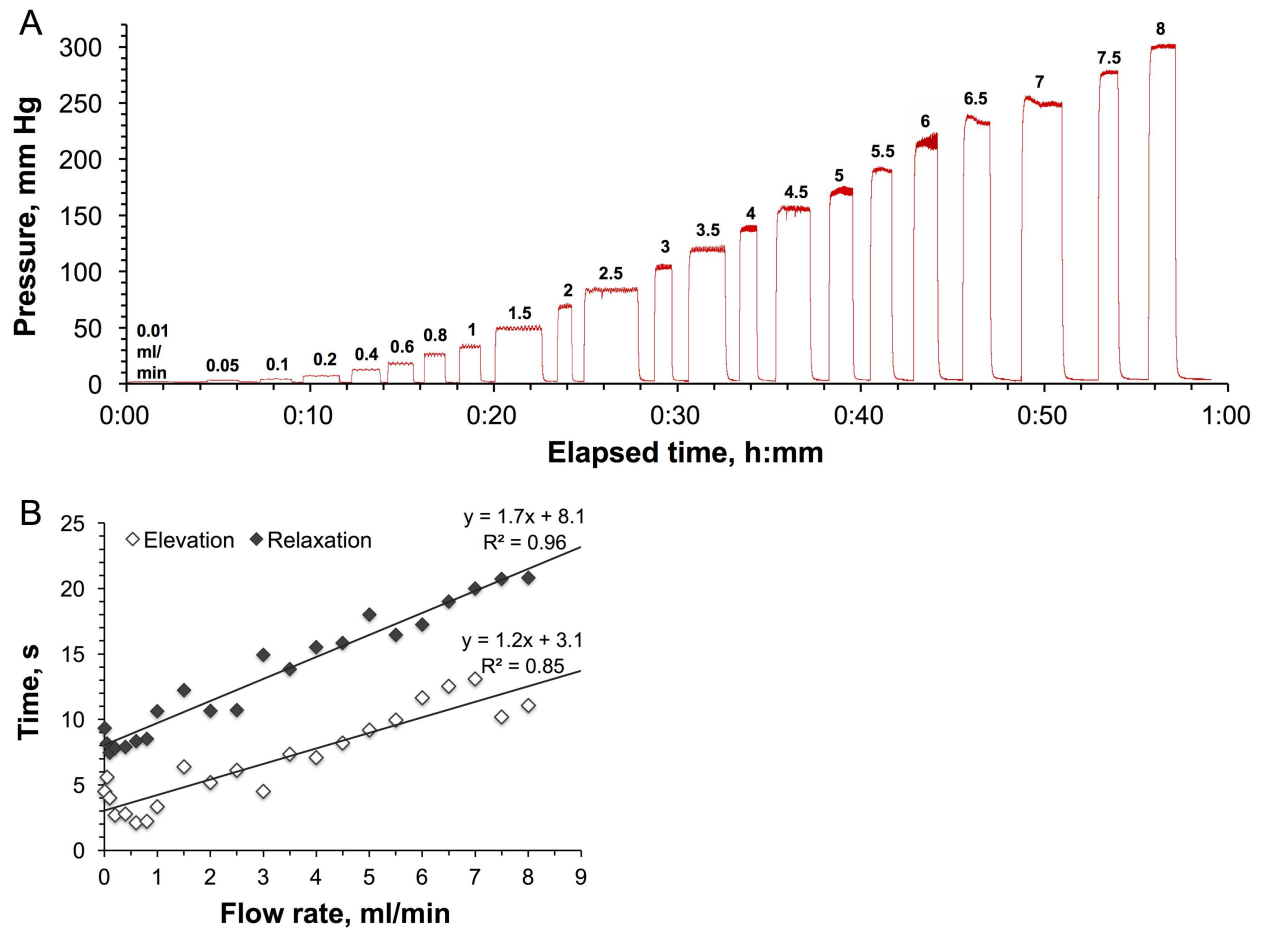

**Figure S5. Characterization of the pressure resistance in the IT injection port in monkeys. (A)** Time dependence of pressure in the IT port measured ex-vivo for various injection flow rates from 0.01 ml/min to 8 ml/min. **(B)** Duration of the pressure elevation and relaxation phases as a function of the injection flow rate.

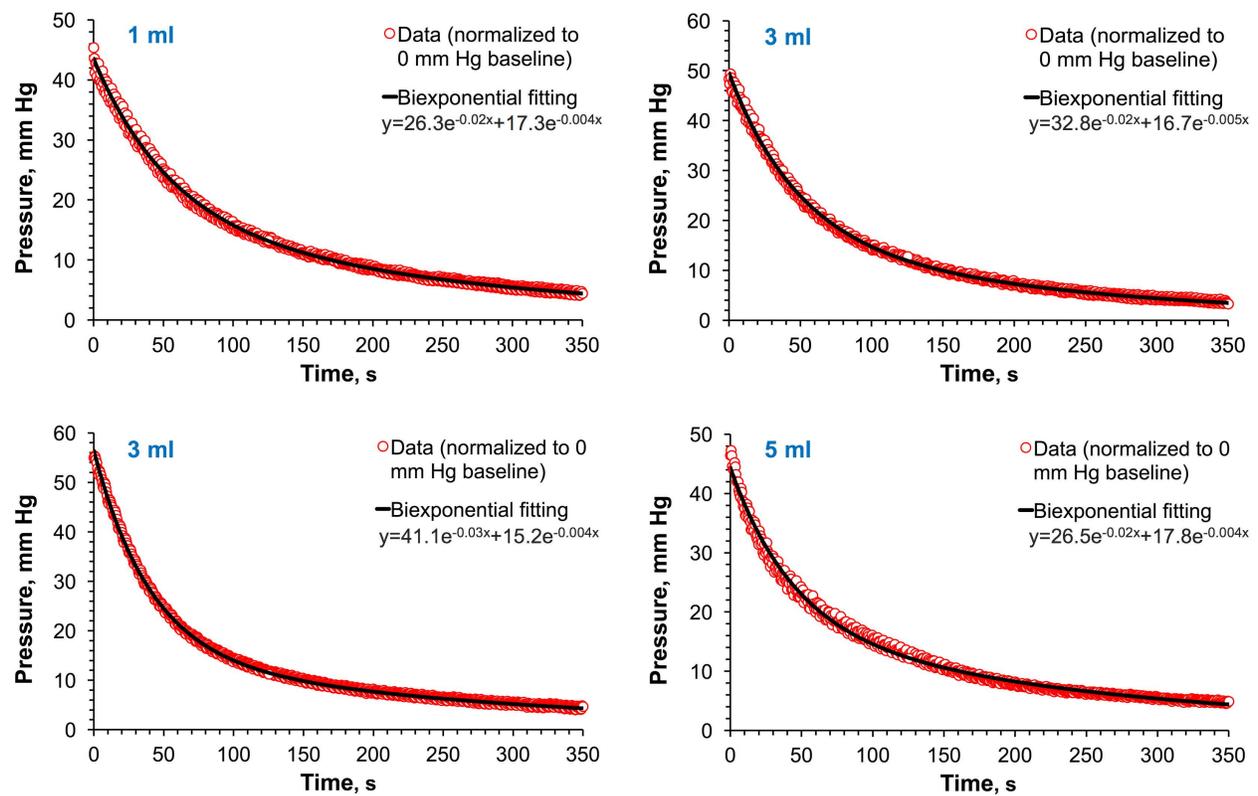

**Figure S6. ICP relaxation to the baseline value after bolus IT injections in the lumbar ports of monkeys.** Left to right: time-dependence curves of ICP (blue dots) and their bi-exponential fitting (red lines) after reaching the peak ICP as a result of administrations at 4.8±1.5, 1.7±0.5, and 2.1 ml/min of 1, 3, and 5 ml of solutes, respectively. All the data are normalized to 0 mm Hg baseline to account for differences in resting pressures.

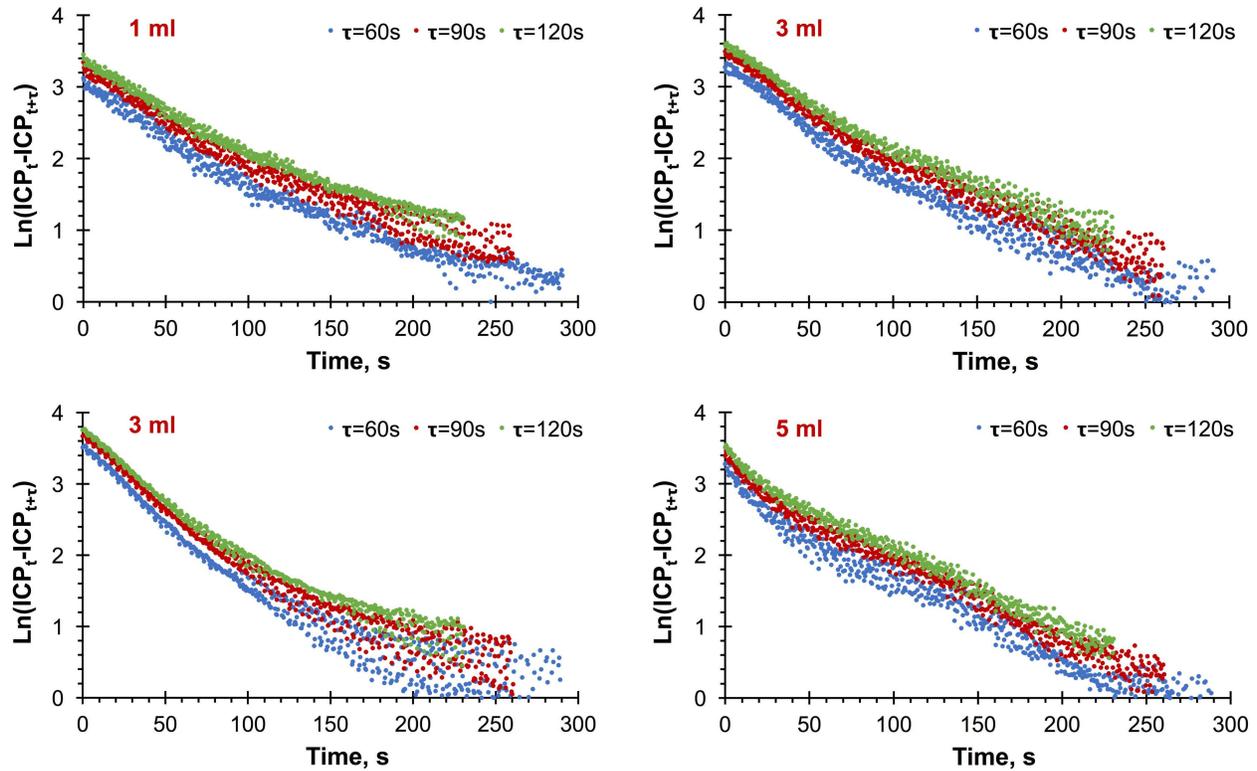

**Fig. S7. Guggenheim linearization of the ICP relaxation after bolus IT injections in the lumbar ports of monkeys.** Left to right: time dependence of the logarithm of difference between two ICP values on the ICP relaxation curves (Fig. S4), which are 60s (blue dots), 90s (red dots), and 120s (orange dots) apart, after administrations at 4.8±1.5, 1.7±0.5, and 2.1 ml/min of 1, 3, and 5 ml of solutes, respectively. A time interval of 300s is shown. Bi-linear shape is indicative of bi-exponential character of the ICP relaxation.

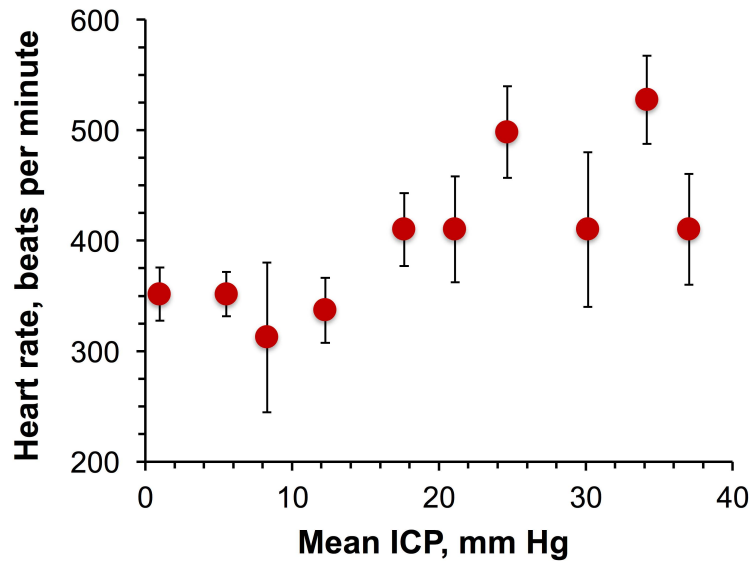

**Fig. S8. Dependence of the heart rate on the intracranial pressure caused in rats for extended durations (>10 min).** The ICP was measured directly in the cisterna magna. The heart rate was obtained by conversion of the P1 frequency component of the ICP pulse wave into beats per minute.

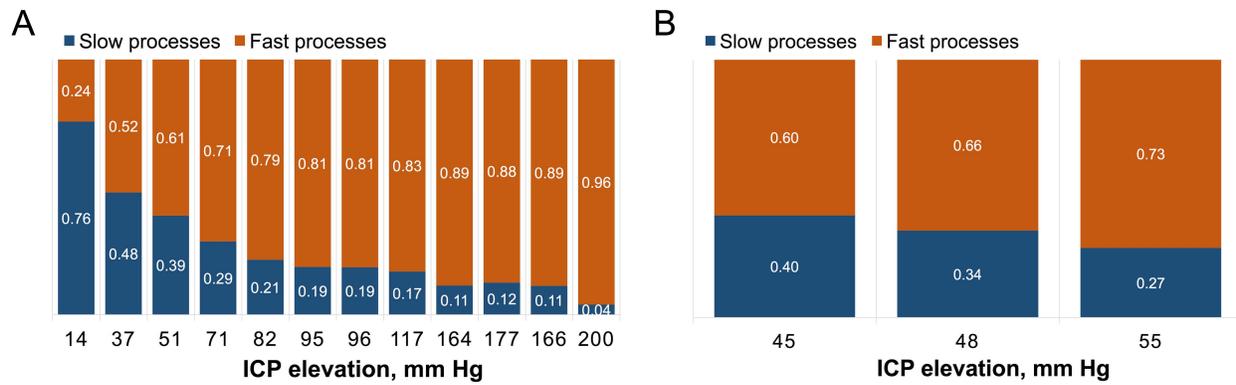

**Fig. S9. Contribution of the slow and fast processes to the ICP relaxation.** The diagrams consist of the relative fractions of the slow (navy) and fast (orange) components of the bi-exponential function used to fit the ICP relaxation curves (Fig. S3 and Fig. S6) after bolus injections in rats (A) and monkeys (B). The fractions were obtained based on the pre-exponential coefficients. The slow and fast components of a bi-exponential function are characterized by longer and shorter half-lives, respectively.